\pdfoutput=1

\documentclass[namedreferences]{solarphysics}

\usepackage[hyperref,optionalrh]{spr-sola-addons} 
\usepackage[pdftex]{graphicx}
\usepackage{color}           
\usepackage{breakurl}        

\ifx \doiurl    \undefined \def \doiurl#1{\href{http://dx.doi.org/#1}{\textsf{DOI}}}\fi
\ifx \adsurl    \undefined \def \adsurl#1{\href{http://adsabs.harvard.edu/abs/#1}{\textsf{ADS}}}\fi
\ifx \arxivurl  \undefined \def \arxivurl#1{\href{http://arxiv.org/abs/#1}{\textsf{arXiv}}}\fi
\ifx \urlurl    \undefined \def \urlurl#1{\href{http://#1}{\textsf{#1}}}\fi
\ifx \mailtourl    \undefined \def \mailtourl#1{\href{mailto:#1}{\textsf{#1}}}\fi

\def\arcdeg{\hbox{$^\circ$}}

\def\gtrsim{\mathrel{\hbox{\rlap{\hbox{\lower4pt\hbox{$\sim$}}}\hbox{$>$}}}}
\def\lesssim{\mathrel{\hbox{\rlap{\hbox{\lower4pt\hbox{$\sim$}}}\hbox{$<$}}}}
\def\sun{\hbox{$\odot$}}

\begin{document}

\begin{article}

\begin{opening}

\title{The Relation Between Large-Scale Coronal Propagating Fronts and Type II Radio Bursts }

\author{Nariaki V.~\surname{Nitta}$^{1}$\sep
        Wei~\surname{Liu}$^{1,2}$\sep
        Nat~\surname{Gopalswamy}$^{3}$\sep
        Seiji~\surname{Yashiro}$^{3,4}$
}
\runningauthor{N.V. Nitta {\it et al.}}
\runningtitle{Relation Between LCPFs and Type II Bursts}

\institute{$^{1}$ Lockheed Martin Solar and Astrophysics Laboratory,
   Department A021S, Building 252, 3251 Hanover Street, Palo Alto, CA
   94304 USA \\
 email: \url{nitta@lmsal.com} \\    
 $^{2}$ W. W. Hansen Experimental Physics Laboratory, Stanford
 University, Stanford, CA 94305 USA  \\
email: \url{weiliu@lmsal.com} \\
 $^{3}$ Code 671, NASA Goddard Space Flight Center, Greenbelt
    20771 USA \\
  email: \url{Nat.Gopalswamy@nasa.gov} \\ 
$^{4}$ The Catholic University of America, Washington, DC
     20064 USA    \\              
email: \url{Seiji.Yashiro@nasa.gov}
             }

\begin{abstract}

Large-scale, wave-like disturbances in extreme-ultraviolet (EUV) and
type II radio bursts are often associated with coronal mass ejections
(CMEs).  Both phenomena may signify shock waves driven by CMEs. Taking
EUV full-disk images at an unprecedented cadence, the {\it Atmospheric
  Imaging Assembly} (AIA) onboard the {\it Solar Dynamics Observatory} has
observed the so-called EIT waves or large-scale coronal propagating
fronts (LCPFs) from their early evolution, which coincides with the
period when most metric type II bursts occur.  This article discusses the
relation of LCPFs as captured by AIA with metric type II bursts.
We show examples of type II bursts without a clear LCPF and fast LCPFs
without a type II burst.  Part of the disconnect between the two
phenomena may be due to the difficulty in identifying them objectively.
Furthermore, it is possible that the individual LCPFs and type II
bursts may reflect different physical processes and external factors.
In particular, the type II bursts that start at low frequencies and
high altitudes tend to accompany an extended arc-shaped feature, which
probably represents the 3D structure of the CME and the shock wave
around it, rather than its near-surface track, which has usually been
identified with EIT waves.  This feature expands and propagates toward
and beyond the limb.  These events may be characterized by stretching
of field lines in the radial direction, and be distinct from other
LCPFs, which may be explained in terms of sudden lateral expansion of
the coronal volume.  Neither LCPFs nor type II bursts by themselves
serve as necessary conditions for coronal shock waves, but these
phenomena may provide useful information on the early evolution of the
shock waves in 3D when both are clearly identified in eruptive events.

\end{abstract}
\keywords{Shock waves; Coronal mass ejections; {\it Solar Dynamics
    Observatory}; Extreme Ultraviolet emission; Radio emission}
\end{opening}


\section{Introduction}
     \label{S-Introduction}

Large-scale wave-like disturbances at coronal
temperatures are one of the most spectacular phenomena revealed by the
{\it Solar and Heliospheric Observatory} (SOHO) \cite{Moses97,Thompson98}.
They were called EUV waves, or EIT waves, after the instrument ({\it
  Extreme-ultraviolet Imaging Telescope} (EIT: \citeauthor{Boudin95},
\citeyear{Boudin95})) that observed them.  \inlinecite{Thompson99}
proposed that EIT waves may be the coronal counterpart of 
chromospheric Moreton--Ramsey
waves observed in H$\alpha$ \cite{Moreton60}, which had been explained
in terms of flare-launched fast-mode magnetohydrodynamic (MHD) shock waves as they were
refracted downward 
\cite{Uchida68}.  The same
shock waves in the corona have been thought to be responsible for type II radio
bursts observed in the metric range 
(\citeauthor{Uchida74}, \citeyear{Uchida74};
\citeauthor{Vrsnak02}, \citeyear{Vrsnak02}, \citeyear{Vrsnak06}).
We therefore
expect a good correlation between EIT waves and type II bursts.
Indeed, \inlinecite{Klassen00} showed that 90\,\% (19 of 21) of their
sample of type II bursts were associated with EIT waves.  
Note that the fast-mode MHD shock waves have become more commonly
attributed to
coronal mass ejections (CMEs) than to flares,
following several studies that showed close relationships with CMEs of both  
EIT waves \cite{Biesecker02,Cliver05,PFChen06}
and type II bursts \cite{Cliver99,Gopalswamy01,Cliver04}.

In contradiction to the interpretation that EIT waves signify fast-mode MHD
waves, many of them were found considerably slower ({\it e.g.}
200\,--\,400~km s$^{-1}$, see \citeauthor{Thompson09}
\citeyear{Thompson09}) than Moreton--Ramsey waves ({\it e.g.}
500\,--\,1200~km s$^{-1}$, see \citeauthor{Warmuth04a},
\citeyear{Warmuth04a}).  Alternatively, some authors linked
EIT waves with direct manifestations of CMEs.
These are called the ``non-wave'' or ``pseudo-wave''
interpretations.  Numerous articles have been published on the nature of
EIT waves (see the latest review by \inlinecite{WeiLiu14} and
references cited therein including several comprehensive reviews).

In retrospect, the controversy over the nature of EIT waves has
been caused largely by inadequate observations of the phenomenon.
In particular, the compromised 10\,--\,20-minute cadence of the EIT
created considerable ambiguities as to 
what should be registered as EIT waves, except for
the small number of canonical or ``typical'' examples 
with a clear front propagating in almost all directions ({\it e.g.} 
\citeauthor{WillsDavey09}, \citeyear{WillsDavey09}).
In the detailed catalog of EIT waves up to June 1998 as compiled by \inlinecite{Thompson09}, many 
events were marked with a low confidence level, which often indicated that
very few images captured the coronal disturbances.  The compromised cadence
was also responsible for the lower speed of EIT waves
(\citeauthor{Long08}, \citeyear{Long08}, \citeyear{Long11}).
Concerning the apparent discrepancy of the
speeds of EIT waves and Moreton--Ramsey waves, \inlinecite{Warmuth04a} showed
in the distance--time plots
that EIT waves could be seen as an extension of Moreton--Ramsey waves, which
experienced deceleration within ten minutes after the onsets.  The EIT
seldom observed these periods.

The problem of the insufficient image cadence has been drastically
ameliorated by the {\it Atmospheric Imaging Assembly} (AIA) onboard 
the {\it Solar Dynamics Observatory}, which was launched
in February 2010. Thanks to the unprecedented 12-second regular cadence 
of EUV full disk images,
we now can trace the large-scale wave-like
disturbances as soon as they are launched.  Here we call them
large-scale coronal propagating fronts (LCPFs) as did
\inlinecite{Nitta13a},
remembering that both fast-mode MHD waves and CME-related features
could contribute to the propagating front
especially during the early evolution (\citeauthor{Schrijver11},
\citeyear{Schrijver11}; \citeauthor{Downs11}, \citeyear{Downs11}).
Shock waves may be present at the flanks of the CMEs more likely 
in the early phase of LCPFs than later, when LCPFs would be identified with
freely propagating MHD waves.  Observations of LCPFs in the early
phase, which usually coincides with the occurrence of the metric type
II burst 
(\citeauthor{Gopalswamy09}, \citeyear{Gopalswamy09}, \citeyear{Gopalswamy13}),
should give us a better insight into how shock waves are driven
in the CME processes.
Many LCPFs in this phase were found to be much faster than
EIT waves, with an average speed of $\approx$\,600~km~s$^{-1}$ \cite{Nitta13a},
although a small number of LCPFs were intrinsically slow ({\it e.g.} 
$<$~300~km~s$^{-1}$).  In contrast to the recent work to group EIT
waves into different origins on the basis of their kinematic behaviors 
\cite{Warmuth11}, 
faster (slower) LCPFs 
do not necessarily decelerate (accelerate) as revealed by 
AIA observations, which contain much faster LCPFs than EIT waves
\cite{Nitta13a}.

The main purpose of the present article is to revisit the
relation of LCPFs with type II bursts.  This is intended as a first step
toward determining when and where shock waves form in solar
eruptions ({\it e.g.}  \citeauthor{Gopalswamy13}, \citeyear{Gopalswamy13}), 
which may have an important consequence in our
understanding of the temporal and spatial properties of solar energetic particle events
({\it e.g.}  \citeauthor{Rouillard12}, \citeyear{Rouillard12}).
To study the association of LCPFs with type II bursts, we examine radio dynamic
spectra rather than relying on the NOAA event lists\footnotemark[1],
which have served as the reference in several statistical studies in
the past.  
In Section 2, we recapitulate the basic results on LCPFs and
their association with type II bursts as presented in the ensemble study by
\inlinecite{Nitta13a}.  
Section 3 gives examples of type II bursts without a clear LCPF and
fast LCPFs without a reported type II burst. This section raises
the question of how objectively the two phenomena can be identified in
the data.  In Section 4 we study the height of the CME at
the type II onset and the starting frequency of the type II burst,
with an emphasis on type II bursts that start at low frequencies and
high altitudes.
Section 5 discusses the implications of our findings, and a summary is
given in Section 6.

\footnotetext[1]{Annual lists of metric radio bursts at
  \url{ftp://ftp.ngdc.noaa.gov/STP/space-weather/solar-data/solar-features/solar-radio/radio-bursts/reports/spectral-listings/}
  until January 2011 and daily event lists thereafter at \url{www.swpc.noaa.gov/ftpmenu/warehouse.html}.}

\section{Overview of the Relation between LCPFs and Type II Bursts} 
\label{Overview}

In \inlinecite{Nitta13a}, a total of 171 LCPFs were manually isolated
in AIA data taken during April 2010\,--\,January 2013. In the absence
of a universal definition of the EIT wave beyond ``the outermost
propagating intensity front reaching global scales''
\cite{Patsourakos12}, our working definition of the LCPF was that
it had an angular expanse of
$\gtrsim$45$\arcdeg$ and propagated at least 200 Mm away from the
center of the associated eruption or flare.  The events are archived at 
\url{aia.lmsal.com/AIA_Waves}, which was created as a side product
of that publication.  This was not meant to be 
a complete list of events, which would be more objectively produced by
automated detection algorithms ({\it e.g.}  \citeauthor{Long14},
\citeyear{Long14}).  Nevertheless, the web site contains extensive movies and
is periodically updated, providing a useful resource to workers on
solar eruptions and related topics.

\begin{figure}    
\centerline{\includegraphics[width=0.999\textwidth,clip=]{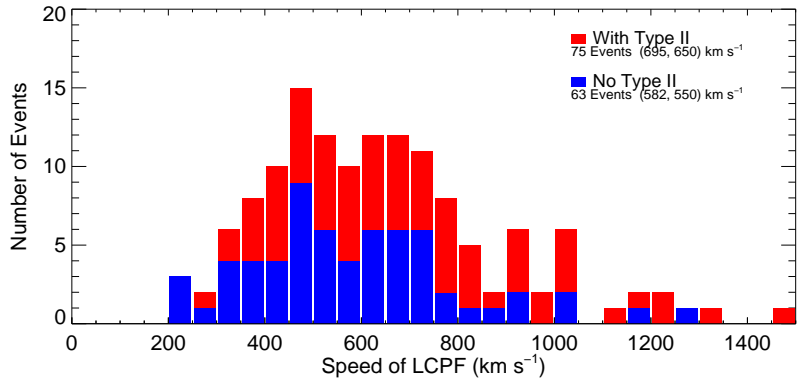}}
              \caption{Distribution of the speed of LCPFs for those associated and not associated with
              a type II burst.  The speeds were measured in AIA
              193~\AA\ images.  The numbers in the parentheses refer
              to average and median speeds.  Adapted from Nitta {\it et al.} (2013a).}
   \label{wave_speed_typeII.pdf}
\end{figure}

As was often the case with past works on EIT waves ({\it e.g.} 
\citeauthor{Thompson09}, \citeyear{Thompson09}), 
\inlinecite{Nitta13a}
concentrated on the events with an arc-shaped front propagating on the solar
disk in AIA images.  There were 138 events in this category.   In other
events, the front was either seen to propagate almost exclusively
along the limb (22 events)
or
too diffuse to trace accurately (11 events).  
All the events propagating along the limb as viewed from Earth 
satisfied the above working definition of the LCPF
in data from the {\it EUV Imager} (EUVI: 
\citeauthor{Wuelser04}, \citeyear{Wuelser04}; \citeauthor{Howard08}, \citeyear{Howard08}) 
on the {\it Solar and Terrestrial Relations Observatory} (STEREO) 
sampled at a 2.5\,--\,5-minute cadence.  
The apparent speed of the front was measured 
from the eruption center along the great circle, averaged  
over the 15$\arcdeg$-wide longitudinal sectors.  This was to take into
account the spherical shape of the surface of the Sun.  The zero height
from the photosphere was assumed.  \inlinecite{WeiLiu14} 
showed that 
the error from this assumption may
be negligible on the disk or if the height of the front does
not significantly change from what has often been cited (see below).
The measurement
was made until the front reached discontinuities such as coronal holes and
active regions, where it was usually deflected ({\it e.g.} 
\citeauthor{Gopalswamy09}, \citeyear{Gopalswamy09}; 
\citeauthor{Olmedo12}, \citeyear{Olmedo12}); it was difficult to trace
the same front after that point.  The sector that both
showed a clear front and yielded the highest speed was recorded, and 
the speed of the front in that sector was entered as the speed of the
event.  For regions close to the limb, the sectors toward the limb
were excluded because foreshortening would seriously impact the
measurement accuracy.  It is also possible that an extended 
arc-shaped feature moving toward and beyond the limb may represent the 3D extent of the
CME and the shock wave around it (see Subsection 3.1 and Section 4).
This is different from what has usually been conceived as an EIT wave, which
is essentially a near-surface phenomenon.  See 
\inlinecite{Patsourakos09} and \inlinecite{Kienreich09} for the height measurement
of the front, yielding the range of
0.1\,--\,0.2~R$_{\sun}$ from the surface.

The distribution of the speed of the 138 LCPFs is given in Figure~1, plotted
separately for those associated (55\,\%) and not associated (45\,\%) with a type II
burst on the basis of the NOAA lists.  We first confirm that many LCPFs are as fast as typical
Moreton--Ramsey waves although a small number of slow ($<$~300~km
s$^{-1}$) fronts exist. 
Additionally, we have not seen super-fast ({\it e.g.}  $>$2000~km
s$^{-1}$) events, which were predicted on the basis of Solar Cycle 23
ground-level-enhancement (GLE) particle events \cite{Nitta12}. This
may be related with the generally weak solar activity of Cycle 24
({\it e.g.}  \citeauthor{Gopalswamy14}, \citeyear{Gopalswamy14}).
Fast LCPFs tend to be associated with a type II
burst but many are not. Conversely, many slow LCPFs are associated with a type
II burst.  Concerning the association in the opposite direction, out
of the 141 type II bursts included in the NOAA lists during the same
period, 85 (60\,\%) were
associated with a clear LCPF including those that propagate primarily
along the limb with negligible
presence on the disk as seen in AIA data.
Most of the remaining type II bursts are associated with eruptive
signatures, if not as extended as LCPFs.

\begin{figure}    
\centerline{\includegraphics[width=0.999\textwidth,clip=]{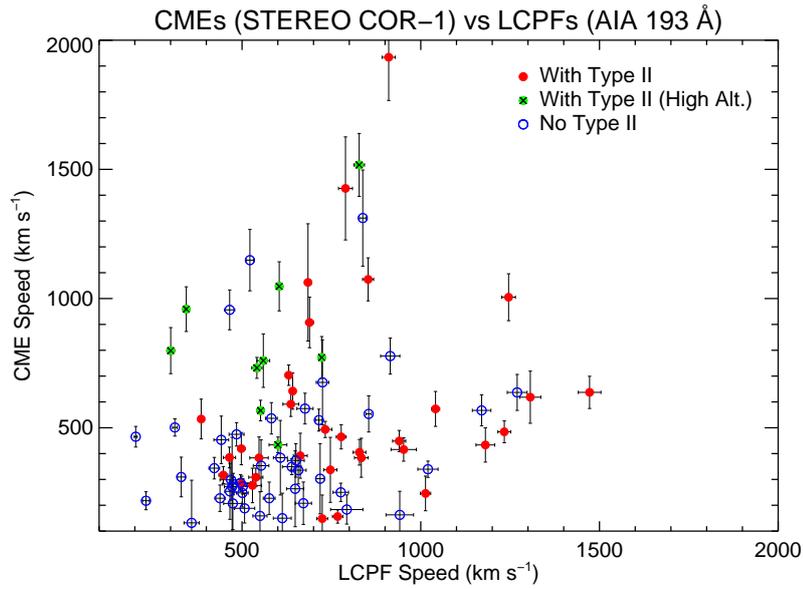}}
              \caption{Speeds of CMEs and LCPFs that were limbward of
                the longitude of 60$\arcdeg$ by STEREO.
This figure corrects Figure~8 in Nitta {\it et al.} (2013a), and also
              distinguishes 
              type II bursts that start at low frequencies and high altitudes (see
              Section 4) 
}
   \label{speeds_lcpf_cme_corrected_encircle_9.pdf}
\end{figure}

LCPFs are often thought to result from sudden lateral expansion of the CME volume
\cite{Patsourakos12}, whereas CMEs are considered to represent expulsion of
coronal structures primarily in the radial direction.
A subset of 86 LCPFs were observed as limb events by either or both of
the twin STEREO spacecraft (``B'' for Behind, and ``A'' for Ahead of the Sun--Earth line), allowing
us to measure the speed of the associated CMEs without serious projection effect.  
Moreover, the STEREO carries the COR-1 coronagraph \cite{Howard08}, 
which observes the corona in the heliocentric distance of (1.4\,--\,4.0)R$_{\sun}$
at a typical cadence of 2.5\,--\,5 minutes.  Therefore the CME kinematic
information calculated from the COR-1 data may be readily
comparable with LCPFs.

Figure~2 shows the CME speed against the LCPF speed for the
86 events, again plotted separately for those associated and not
associated with a type II burst. This plot suggests the existence of
two classes of events, depending on whether the lateral expansion
(LCPF) is faster than the radial expansion (CME).  
Type II bursts may be associated with LCPFs irrespective of which
speed is higher.  
As was the case with LCPFs, fast CMEs tend to be associated with a type II burst, but again
several of them are not. A handful of type II bursts are observed when both
the associated LCPF and CME are slow ({\it e.g.}  $<$\,600~km~s$^{-1}$).
\inlinecite{Nitta13a} suggested that neither $v_{\tt LCPF}$ nor
$v_{\tt CME}$  
may be a deciding factor for the LCPF to be associated
with a type II burst.

\section{Revisiting the Relation between LCPFs and Type II Bursts} 
      \label{Main section}

Type II bursts, identified with slowly drifting narrow-band features in radio dynamic spectra, 
have long been known as a good indicator of MHD shock waves produced
by solar eruptions
({\it e.g.}  \citeauthor{Nelson85}, \citeyear{Nelson85};
\citeauthor{Gopalswamy00}, \citeyear{Gopalswamy00}).
Previous statistical studies of type II bursts often relied on the NOAA lists, 
but we choose to examine the dynamic spectra and to understand
what they objectively tell us.  
Unlike modern spacecraft data of the Sun, radio dynamic spectra produced by
ground-based observatories had not been
easily accessible apart from pre-scaled graphic files.  
However, the dynamic spectra  
obtained at the US Air Force {\it Radio Solar Telescope Network} (RSTN)
sites have been archived as raw data and made available through the National Geophysical Data
Center since the
early 2000s. The daily files are presently downloadable at
\url{www.ngdc.noaa.gov/stp/space-weather/solar-data/solar-features/solar-radio/rstn-spectral/}.
Note that the archive is not perfect, and data of some well-known type
II bursts are missing.  In addition, some useful information on the
RSTN dynamic spectra
is available in \url{ftp://ftp.ngdc.noaa.gov/STP/space-weather/solar-data/solar-features/solar-radio/rstn-spectral/docs/SRSData.ppt}.

We study the RSTN dynamic spectra for
almost all the type II bursts included in the NOAA lists during April
2010\,--\,January 2013, and for all the LCPFs in the same period as analyzed by 
\inlinecite{Nitta13a}.  Although the frequency range of the RSTN dynamic spectra
is limited to 180\,--\,25~MHz, there is a unique advantage of
around-the-clock coverage,
which is made possible
by the four RSTN sites in Canada,
Australia, Hawaii, and Italy.  Many type II bursts are observed at more
than one RSTN site, allowing us to evaluate the consistency of the
spectra.

\begin{figure}    
\centerline{\includegraphics[width=0.999\textwidth,clip=]{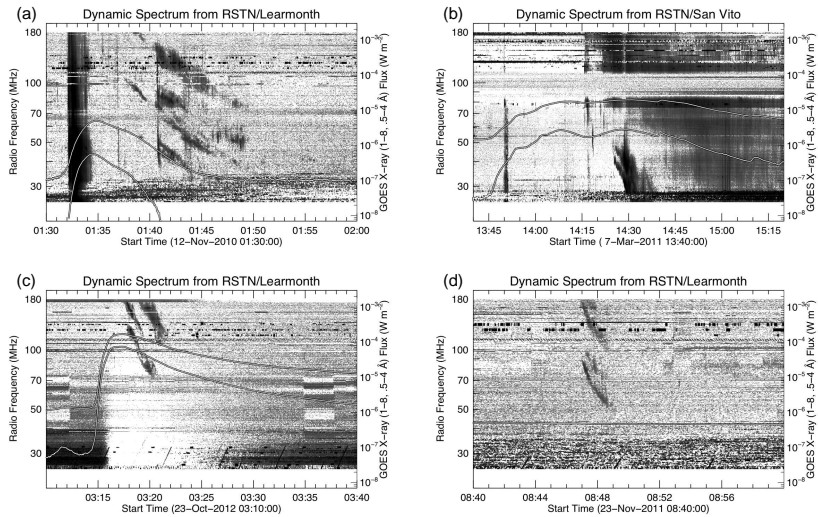}}
              \caption{Radio dynamic spectra of four type II bursts
                that are not associated with a clear LCPF.  The GOES
                X-ray light curves are also plotted in each dynamic
                spectrum. The associated flares are: (a)
                SOL2010-11-12T01:34\protect\footnotemark[2] (C4.6,
                S24E03), 
(b) SOL2011-03-07T14:30 (M1.9, N21E12), (c) SOL2012-10-23T03:17
                (X1.8, S12E58) and (d) None.
}
   \label{rstn_spectra_4typical.pdf}
\end{figure}

\footnotetext[2]{For SOL identification convention, see {\it Solar
    Phys.} {\bf 263}, pp.1 – 2, 2010.}

As shown in Subsection 3.2 and Section 4, it is not always easy to find a type
II burst objectively in the dynamic spectrum, partly because of other bursts or non-solar features
including interference.  
Furthermore, the real dynamic
spectra rarely show clear and isolated lanes of fundamental and
harmonic emissions.  Indeed, we could not confidently confirm all the type
II bursts in the NOAA lists, and yet it was also difficult to rule out the
slowly drifting features possibly undermined by the ongoing type III or IV bursts.  In
other words, we accept all the type II bursts in the NOAA lists, except for a
few cases that correspond to no radio bursts of any kind, because of 
obvious typographic errors. 
Certain type II bursts may show greater presence above 200 MHz or below 30 MHz.
They may be left out in the NOAA lists
(S. M. White, 2013, 2014, private communication).  
Here we have not taken extra steps to
find additional type II bursts outside the RSTN spectra, but when there
is ambiguity about the presence of a type II burst in the RSTN data we
access the radio dynamic spectra
as provided by other observatories, including Greenbank and BIRS
(\url{www.astro.umd.edu/~white/gb/index.shtml}), Nan\c{c}ay
(\url{secchirh.obspm.fr/select.php}), Hiraiso
(\url{sunbase.nict.go.jp/solar/denpa/index.html}), IZMIRAN
(\url{www.izmiran.ru/stp/lars/}), and e-Callisto (\url{e-callisto.org/}).

Because our immediate goal is to find the trend and range of the
relation between LCPFs and type II bursts, study of sub-features of type
II bursts such as herringbones and band splits 
({\it e.g.}  \citeauthor{Nelson85}, \citeyear{Nelson85}; 
\citeauthor{Carley13}, \citeyear{Carley13}), although
important, is outside the scope
of this work.  We also do not make use of radio imaging
observations.  During the period of interest, they have been limited to
the frequencies above 150~MHz, and there is almost no direct overlap with the
type II bursts observed at RSTN.

\subsection{Type II Bursts Not Associated With a Clear LCPF}

Figure~3 shows dynamic spectra of four
events in this category.  In this figure as well as Figures 6 
(in Subsection 3.2) and 8 (in Seciton 4),
the spectra are displayed in negative with a fixed scaling that 
we found was good to bring
up weak signals. 
The GOES X-ray light curves are also
included in the plots. In Figure~3a we note a clear type II burst
consisting of four lanes that correspond to two fundamental--harmonic
pairs. They are preceded by strong type III bursts
during the rising phase of a short-lived C-class flare.  In
Figure~3b, the type II burst may be only barely seen at low
frequencies between
14:25 and 14:35 UT; the dynamic spectrum is dominated by a
broad-band type IV burst.  Most features in the dynamic spectrum
started with the second episode of the long-duration M1.9 flare.
The type II burst in Figure~3c is associated with an X1.8 flare.  
It is band-split, and the starting frequency in
the upper fundamental track seems to be above the RSTN range. 
In Figure~3d the type II burst is very short-lived and no X-ray flare
is registered as the GOES X-ray light curves remain flat.

\begin{figure}    
\centerline{\includegraphics[width=0.999\textwidth,clip=]{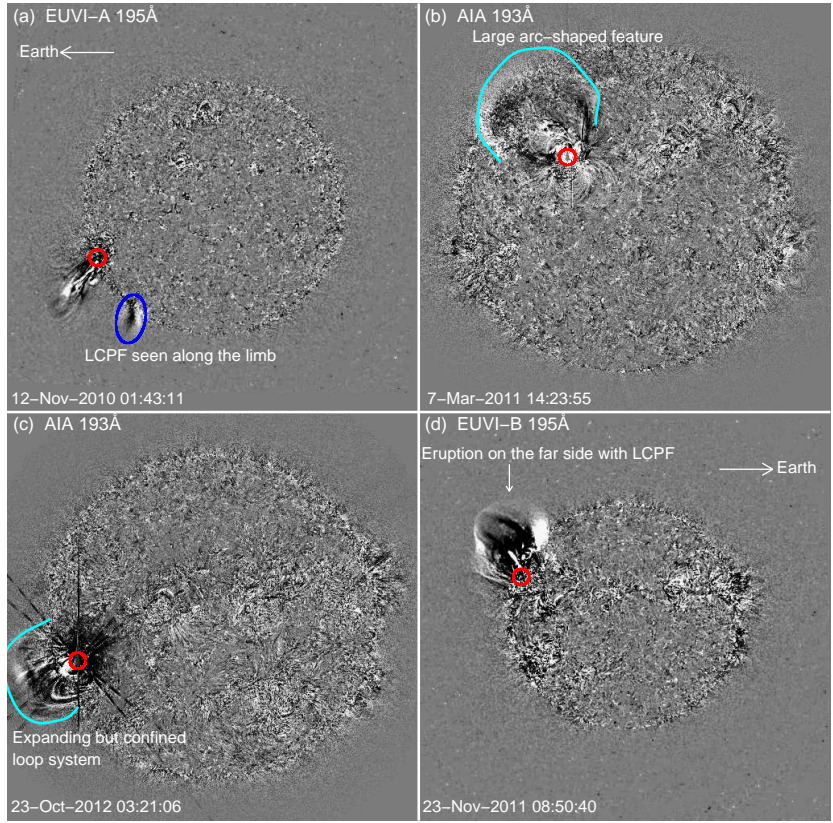}}
              \caption{EUV difference images of the four events in Figure~3. 
 The associated flares are encircled in red
                and some features presumably related with type II
                bursts are also annotated.
}
   \label{images4typeII.pdf}
\end{figure}

In Figures~4a\,--\,d we show solar images around the times of the four type II bursts
in Figures~3a\,--\,d.  They are EUV difference images from the AIA (193~\AA)
and EUVI (195~\AA). The temporal differences of these images are
144~seconds for AIA and 150 or 300 seconds for EUVI.
The associated flares are encircled in red.

The type II burst in Figure~3a is associated with a minor filament
eruption but not with a clear LCPF in AIA images.
We show in Figure~4a an image from the EUVI
on STEREO-A (EUVI-A), which gives a limb view
of the event.  At the 150-second cadence, EUVI-A images captured a
propagating front along the limb (counterclockwise). Encircled
in blue is the feature that corresponds to the front during the type
II burst.  This is an example that indicates that a LCPF can be more
difficult to observe on the disk than along the limb.

The image in Figure 4b is dominated by a large arc-shaped feature (traced
in cyan), which is seen to expand and to propagate toward and beyond the northeast
limb.  This likely represents the 3D structure of the CME
and the shock wave around it, rather than its near-surface track,
which has usually been linked with an EIT wave.  This event was
actually included in \inlinecite{Nitta13a}, because there was a diffuse
front that could be traced toward north before the arc-shaped feature
became prominent, which coincided with the onset of the type II burst. 
The speed of the diffuse front was only $\approx$\,300~km~s$^{-1}$. 
The difficulty of tracing the diffuse front, especially after the
arc-shaped feature started to dominate, puts this event in the ``no clear
LCPF'' category.  In Section 4, we show more events that are similar.

In Figure~4c the expanding loop
system is indicated in cyan, but unlike the arc-shaped feature
in Figure~4b, it becomes too diffuse to follow at a certain distance.  Data from the
EUVI-B, from which the flare was W63, show the same
pattern.  This event may therefore be a confined ejection.  As
expected, there was no associated CME.  
Moreover, there was no front propagating along the limb in either AIA
or EUVI-B data.  This is another
example of a type II burst that starts at a high frequency and has no
associated CME ({\it cf.} \citeauthor{Magdalenic12}, \citeyear{Magdalenic12}).

Figure~4d is an EUVI 195~\AA\ difference image taken just after 
the short-lived type II burst in Figure~3d.  It is clear that the
type II burst was associated with an eruption on the far side as
was captured in EUVI-B images at the 300-second cadence. 
Based on the EUVI 195~\AA\ flux, the eruption was likely associated with an
M-class flare if observed by the GOES {\it X-ray Sensor} (XRS) (see
\citeauthor{Nitta13b}, \citeyear{Nitta13b} for the EUVI--XRS flux relation).
The type II burst was seen only when 
the associated shock wave became visible from Earth and was located at the right range
of density for the RSTN frequencies.  

\begin{figure}    
\centerline{\includegraphics[width=0.999\textwidth,clip=]{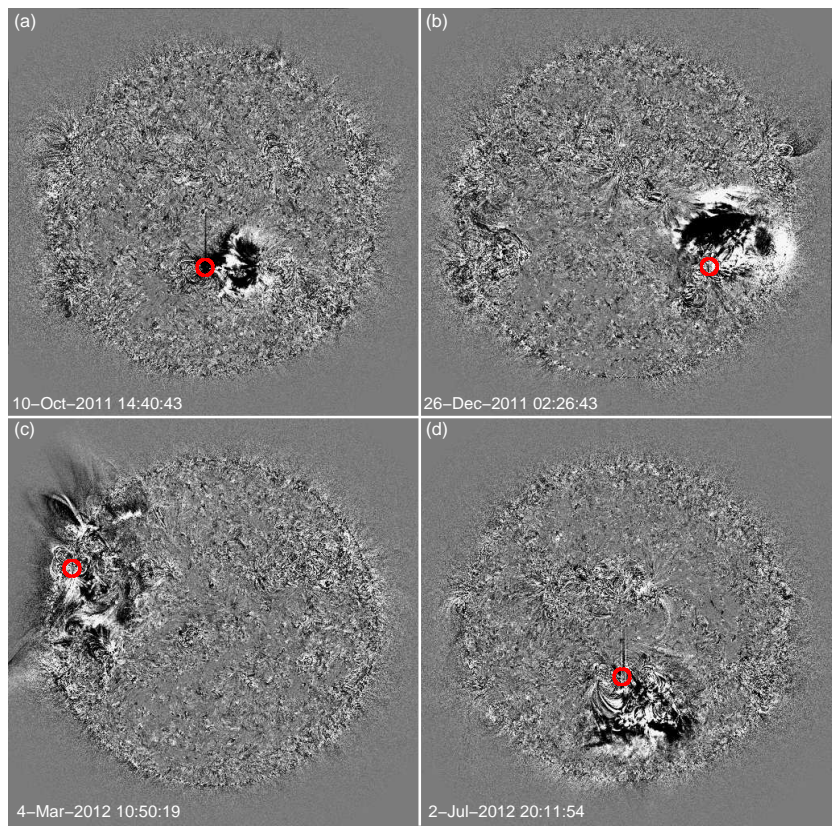}}
              \caption{AIA 193~\AA\ running difference images of four
                fast LCPFs not associated with a clear type II burst.
                The associated flares are encircled in red.  
}
   \label{images4notypeII.pdf}
\end{figure}

\subsection{Fast LCPFs Not Associated With a Type II Burst}

We tend to link faster LCPFs with shock waves and hence expect a
more frequent association with type II bursts.  
Indeed, such a trend is seen in Figure~1, but it is not strong.  
For example, 8 LCPFs are faster than 800~km s$^{-1}$ and not associated with a
type II burst, according to the NOAA lists.  
In Figure~5, we show AIA 193~\AA\ difference images of four of them.  

The LCPF in Figure~5a was associated with a short-duration C-class flare, and
with a marginal CME in COR-1 data viewed as a limb event. Nevertheless, it was
a clear LCPF up to 320~Mm from the flare center.
The LCPF in Figure~5b was produced by 
a more energetic eruption which was associated with a wide (angular width
$\approx$\,130$\arcdeg$) CME as observed by 
the {\it Large Angle Spectroscopic Coronagraph} (LASCO) on SOHO.
The LCPF in Figure~5c, occurring close to the limb,
looked even more spectacular.
In addition to the slow ($\approx$\,300~km~s$^{-1}$) fronts along the limb,
both clockwise and counterclockwise, a fast front was seen to
propagate on the
disk.  The eruption that produced this LCPF 
was associated a fast ($\approx$\,1300~km~s$^{-1}$)
full halo CME observed by LASCO.  It was also accompanied by quasi-periodic
fast propagating (QFP) wave trains ({\it e.g.}  \citeauthor{WeiLiu12},
\citeyear{WeiLiu12}),
which may be one of the signatures to indicate more energetic eruptions.
 The LCPF in Figure~5d came from an extensive filament
eruption that resulted in the reformation of a large area in the south polar region.  It was
associated with a wide ($\approx$\,145$\arcdeg$) CME.

\begin{figure}    
\centerline{\includegraphics[width=0.999\textwidth,clip=]{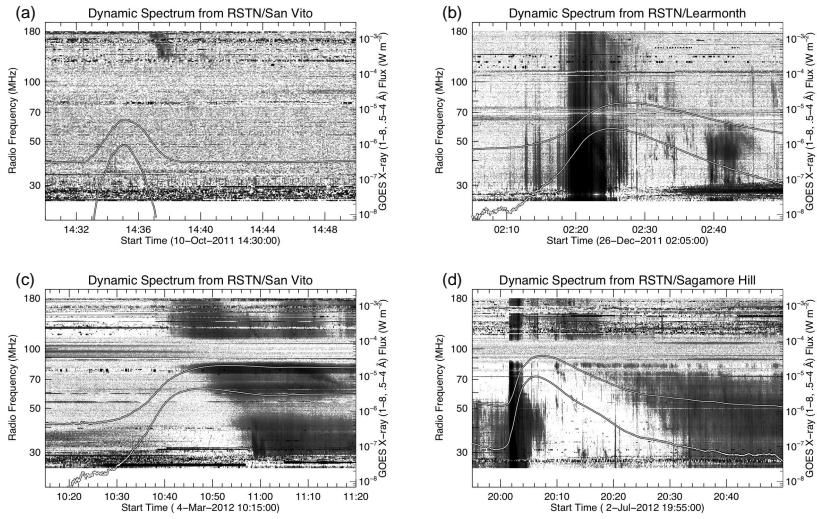}}
              \caption{Dynamic spectra of radio bursts associated with
                the four events shown in Figure~5.  As in Figure~3,
                the GOES X-ray light curves are also shown.  The associated
                flares are: (a)
                SOL2011-10-10T14:35 (C4.5, S13E03), (b)
                SOL2011-12-26T02:27 (M1.5, S21W33),
                (c) SOL2012-03-04T10:52 (M2.0, N17E64) and (d)
                SOL2012-07-02T20:07 (M3.8, S17E03).  No type II bursts were
                reported in these periods.
}
   \label{rstn_spectra_notypeII.pdf}
\end{figure}

The dynamic spectra and X-ray light curves of 
the four events in Figure~5 are shown in
Figure~6.  The dynamic spectrum of Figure~6a shows almost nothing, but we note
a short-lived (14:37\,--\,14:38) feature above 130 MHz.
This may be a small type II burst.  The spectra of
Figures~6b\,--\,d are dominated by other types of bursts: type III burst in
Figure~6b, type IV burst in Figure~6c 
and both type III and type IV bursts in
Figure~6d.  It is difficult to find a slowly drifting narrow-band feature
in these spectra, but we also cannot convincingly rule out the possibility that it
may exist behind 
the dominant features.  
Complex spectra of this kind are characteristic of energetic eruptions
that are associated with GLEs \cite{Gopalswamy12b,Nitta12}.
Accordingly, we may not be able to conclude that 
any of these four examples lack a type II burst.  In fact,
the confidence level of a type II burst may not be much higher
in many energetic events that are associated with a type II
burst as included in the NOAA lists.

\section{CME Height at the Onset of Type II Bursts and Different
  Kinds of LCPFs}

In this section, we study  where the shock wave forms by
measuring the height of the CME when the type II
burst starts.  We may safely assume that the stand-off distance of the
shock wave from the CME is still small at its genesis. 
\inlinecite{Gopalswamy13} used two methods to study the height of shock
formation, namely i) direct measurement of the CME viewed near the limb (to minimize
projection effect), and ii) the so-called wave diameter method, where the outer edge
of the LCPF on the disk was fitted to a circle and the radius of the circle was equated with
the height of the CME shock, assuming spherical geometry.  Here we
exclusively use method i) as applied to data from STEREO (EUVI
and COR-1),
which observed the CMEs as limb events (see Figure~2).  Out of the 86 LCPFs included in
Figure~2, 42 are associated with a type II burst as included in the
NOAA lists.
A further advantage of this method is an overlap of fields of view
of the EUVI and COR-1, which is not the case with the AIA and
LASCO.
Moreover, we can study the dynamic evolution of the LCPFs on the
solar disk better with AIA than with EUVI because of the higher cadence.

\begin{figure}    
\centerline{\includegraphics[width=0.999\textwidth,clip=]{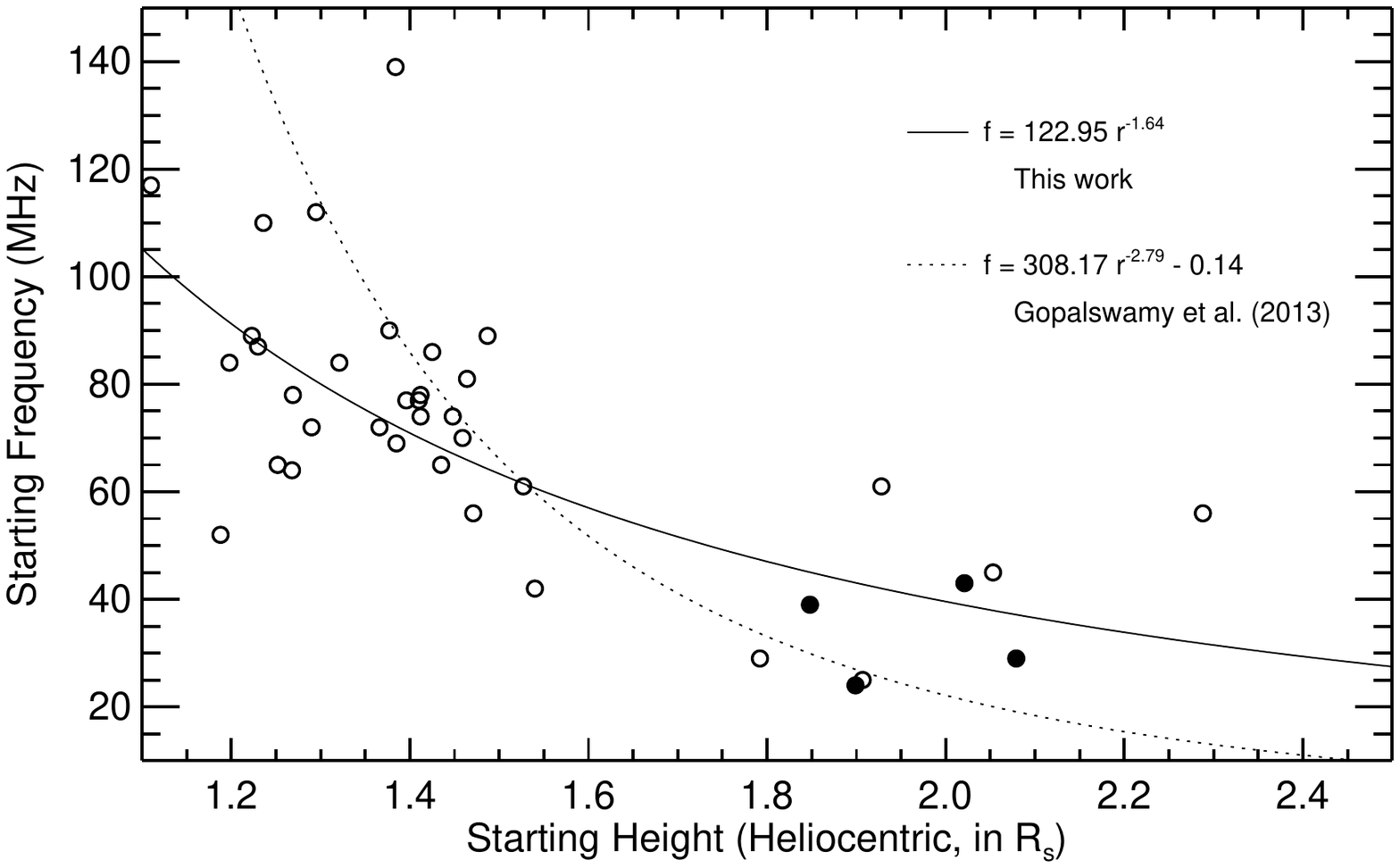}}
              \caption{CME height vs starting frequency of the type II
                burst for 42 events that were observed as limb events
                by STEREO.  The four events that appear in Figures 8
                and 9 are plotted with filled circles.
}
   \label{typeII_height_freq.pdf}
\end{figure}

We expect the shock formation height to inversely correlate with the
starting frequency of the type II burst, reflecting the decreasing
density of the solar corona with height.  
It appears that the frequency range shown in
the NOAA lists is sometimes different from what we see in the dynamic spectrum.
Therefore we re-evaluate the RSTN spectra to obtain independent 
ranges of time and frequency (fundamental) of the type II bursts, 
as in
Table~1 for the four events discussed below.
We assume that a slowly drifting feature in the dynamic spectrum corresponds to the harmonic
if no emission is observed at about twice the frequency.  As a result,
our frequency range tends to be lower than that reported in the
NOAA lists, especially when the latter does not cover the full RSTN
range (180\,--\,25~MHz). 

Figure~7 plots the starting frequency against the starting height of type II bursts.
As expected, we immediately note a trend that type II bursts that start at low
altitudes have high starting frequencies. 
Many start below the height (heliocentric) of
1.5~R$_{\sun}$, and 
a smaller number of type II bursts start around 2~R$_{\sun}$.  
This is basically consistent with the recent results by
\inlinecite{Gopalswamy13}. The difference as reflected in the fits
may partly come from their primary use of 
the wave diameter method and their inclusion of events with high
starting frequencies; here we exclude the 13 February 2013 event,
whose starting frequency was well above the RSTN range.
In Figure~7 we count nine type II bursts whose starting height are
$>$1.7~R$_{\sun}$  and starting frequencies $<$60 ~MHz.  
The plot suggests that they may represent 
a different class of events reflecting different
physical processes and external conditions.
The event shown in Figures~3b and 4b is one
of them.

\begin{figure}    
\centerline{\includegraphics[width=0.999\textwidth,clip=]{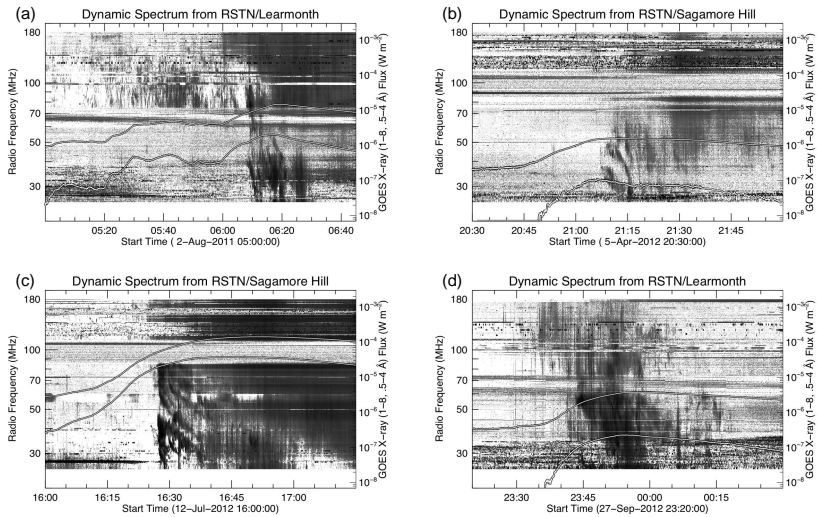}}
              \caption{Dynamic spectra of four type II bursts that
                start at low frequencies and high altitudes.  The associated flares are:
                (a) SOL2011-08-02T06:19 (M1.4, N15W14), (b)
                SOL2012-04-05T21:10 (C1.5, N17W32),
                (c) SOL2012-07-12T16:49 (X1.4, S14W02) and (d)
                SOL2012-09-27T23:57 (C3.7, N09W33).
}
   \label{rstn_spectra_typeII_low.pdf}
\end{figure}

\begin{table}

\caption{Reported and revised time and frequency ranges of the type II bursts shown in Figure~8.
}
\label{from_Event.txt}

\tabcolsep 5.8pt
\begin{tabular}{cccccc}     
  \hline                   
  & Range of$^{1}$ & & Range of$^{1}$ & Range of$^{2}$ & Range of$^{2}$ \\
Date & Time & Station & Frequency & Time &
Frequency \\ 
\hline
 2 Aug 2011 & 06:08\,--\,06:22  & CUL & 90\,--\,23  & 06:08\,--\,06:26
 & 39\,--\,15 \\
 5 Apr 2012 & 21:08\,--\,21:16  & PAL & 43\,--\,25  & 21:08\,--\,21:15
 & 23\,--\,19 \\
12 Jul 2012 & 16:25\,--\,16:53  & SAG & 82\,--\,25  & 16:28\,--\,16:40
 & 43\,--\,25 \\
27 Sep 2012 & 23:44\,--\,23:54  & CUL & 60\,--\,27  & 23:43\,--\,23:53
 & 29\,--\,17 \\
\hline
\end{tabular}

\noindent
1: From the NOAA lists.
2: From our analysis.

\end{table}

We discuss four of these nine events, which are indicated by filled
circles in Figure~7.  Figure~8 shows their dynamic spectra with soft
X-ray light curves.  We note that none of them are
short-duration flares.  
As in some of the spectra shown in Figures~3 and 6, the type II-like
feature may not be objectively located.  Nonetheless, Table~1
indicated that we choose essentially the same intervals as the RSTN
observers.  The difference in the frequency range may likely come from
whether the given feature is interpreted to be either a fundamental or 
harmonic emission.

\begin{figure}    
\centerline{\includegraphics[width=0.999\textwidth,clip=]{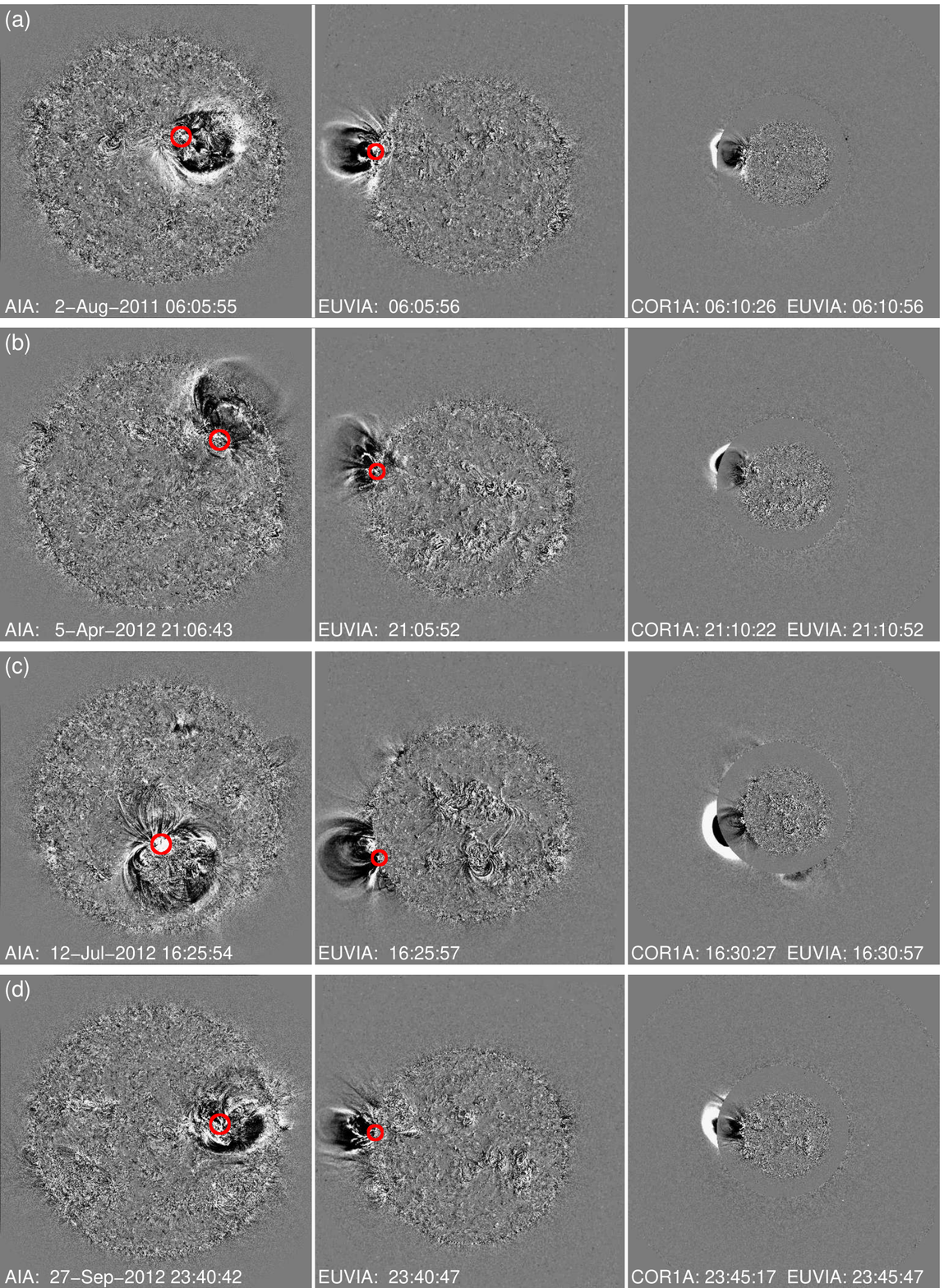}}
              \caption{Images of the four events in Figure~8.
The left column gives AIA images within a few minutes before the onset
         of type II bursts.  The middle column shows limb views of the
         events in EUVI images around the times of the AIA images in
         the left panels.  The associated flares are encircled in red
         in the left and middle panels.  The right column shows COR-1 images just
         after the onset of the type II bursts.  They show CMEs
         already at $\gtrsim$2~R$_{\sun}$ from disk center.}
   \label{images4typeIIlow.pdf}
\end{figure}

In Figure~9, we show images taken around the onsets of these four
events.  
As in Figure~4b, a large arc-shaped feature is present in each
event, and it is seen in movies 
to expand and to propagate toward and beyond the limb.
This indicates that these features
may not be 
the near-surface track of the coronal shock wave. Instead, they are
probably the 3D structure of the CME and the shock wave around it.  The middle and right
panels give the limb view of these fronts, indicating that the
arc-shaped feature on the disk corresponds to the height structure of
the CME.
In this type of event, 
it seems that the radial motions dominate over
the lateral motions.  
In Figure~2, we set aside these nine type II bursts that start at
low frequencies and high altitudes. They tend to have
$v_{\tt CME} > v_{\tt LCPF}$.
However, other events also fall in this category.
The lack of clear grouping may partly come from inadequate
measurements of the speeds of the CMEs and LCPFs. They are average
speeds, not capturing, for example, the impulsive acceleration of CMEs
close to the Sun ({\it e.g.} \citeauthor{Bein11} \citeyear{Bein11}).
In addition, the LCPF generally precedes the CME,
meaning that the speeds are not measured simultaneously.

\section{Discussion}

The initial motivation of this work was to understand the reason for
the modest (50\,--\,60\,\%) association of LCPFs with type II bursts
\cite{Nitta13a}.  In particular, we
ask why the association does not depend strongly on $v_{\tt LCPF}$.
Are there different types of LCPFs and type II bursts?  To complicate
the issue, we first point out that the association of LCPFs 
with type II bursts may not only
depend on their properties.  It may also depend on the presence of a
previous CME, which can have the effect of lowering the
fast-mode speed of the corona in which the next CME is injected.
In such a ``preconditioned'' corona, a shock wave can
more easily form for the same driver (CME) speed \cite{Yashiro14}.  
This is an interesting idea that needs to be validated 
using a larger sample of events.

We note that past statistical studies of type II bursts had often
relied on the NOAA lists and the information contained therein.  On
the other hand, the RSTN dynamic spectra have been available as raw
data since the early 2000s, making it possible for us to analyze them
and to extract more information than given in the NOAA lists.  We
immediately realize that type II bursts may not be objectively
identified in dynamic spectra. We rarely find a clear
fundamental-harmonic emission pair, and type III and IV bursts may
occur around the same time especially in large events.  Even when we
have a clear type II burst, its association with a LCPF depends on
the frequency range in which it is observed.  
Even within the limited RSTN frequency range,
those start at higher frequencies may not involve a LCPF or CME ({\it e.g.} 
the event shown in Figures~3c and 4c), similar to the type II
bursts studied by \inlinecite{Magdalenic12}.  There are also a number
of type II bursts that start below 30~MHz and are not included in the
NOAA lists.  They are often not associated with a CME, flare, or LCPF.

We experience similar difficulties with identifying LCPFs.  We look in
AIA data for signatures similar to what researchers have considered as
EIT waves, namely an arc-shaped front propagating a large distance and
in a wide angular extent.  This excludes smaller events ({\it e.g.} 
\citeauthor{RZheng11}, \citeyear{RZheng11}), but the distinction could
be arbitrary.  Moreover, LCPFs are sometimes too diffuse to detect on
the disk.  The same event can be better seen to propagate along the
limb if viewed from the side (or close to the limb) as is the case for
the event given in Figures~3a and 4a.
Some LCPFs that are seen to propagate along the limb also manifest on
the disk ({\it e.g.} the event in Figures~5c and 6c).  They are probably
more energetic events with greater East--West extensions.

AIA has revealed the propagation of an arc-shaped feature from the
solar disk toward and beyond the limb, which should be considered to
be the 3D structure of the CME and shock wave rather than only their
near-surface tracks.  This pattern appears to be characteristic of
type II bursts that start at low frequencies and high altitudes.  The
height of shock formation should closely reflect 
the height dependence of the CME speed and the fast-mode speed.
The large arc-shaped feature in these events seems to expand more radially
than laterally, and the associated near-surface feature is slow and diffuse.
This reminds us of the scenario proposed by \inlinecite{PFChen02}, in
which EIT waves as near-surface phenomena are dictated by stretching
of field lines in the radial direction, resulting in $v_{\tt CME} >
v_{\tt LCPF}$ \cite{PFChen02}.  The concept is aligned with the
so-called standard model of eruptive flares due to magnetic reconnection in 2D
(or 2.5D) settings.  However, this type of LCPFs is a minority and other
LCPFs seem to be driven by lateral expansion \cite{Patsourakos12}, 
which may not be naturally incorporated in
the standard model.  Note that the standard model has occasionally
been applied to conflicting or at least ambiguous observations 
({\it e.g.} \citeauthor{Nitta10}, \citeyear{Nitta10}), a tendency that
needs to be corrected in advanced observations.
It is possible that these ``fat'' CMEs due to lateral expansion may
need different mechanisms from those that are already part of the standard
model and applicable to ``tall'' CMEs dominated by radial expansion.


From the perspective of studying coronal shock waves due to solar
eruptions, it is more productive to consider the 3D structure of the
CME rather than to stick to the strict definition of LCPFs or EIT
waves as the manifestations close to the surface.  Following an
example from EIT \cite{PFChen09}, a dome-like feature was observed by
EUVI \cite{Veronig10}.  A shock wave likely forms ahead of the CME,
which more often appears off the limb (see the 13 June 2010 event
studied by \citeauthor{Kozarev11}, \citeyear{Kozarev11};
\citeauthor{Ma11}, \citeyear{Ma11}; \citeauthor{Gopalswamy12a},
\citeyear{Gopalswamy12a}).  The shock wave initially propagated 
faster at the nose
than at the flanks, which corresponds to what were conceived as EIT
waves.  The observed kinematic behaviors have been partially
reproduced in theory and models (\citeauthor{XHZhao11},
\citeyear{XHZhao11}; \citeauthor{Grechnev11}, \citeyear{Grechnev11};
\citeauthor{Temmer13}, \citeyear{Temmer13}) of the 17 January 2010
event reported by \inlinecite{Veronig10}.  For the 13 June 2010 event,
\inlinecite{Kouloumvakos14} concluded that both the nose and flank of
the CME drove the shock.  In a major eruptive event on 22 September
2011, \inlinecite{Carley13} presented radio-imaging
observations that located the shock wave on one of the flanks.  Future
observations may test whether the shock wave is driven primarily in the
radial direction for type II bursts that start at low frequencies and
high altitudes.

Although every type II burst is associated with a shock wave,
not all of the shocks manifest as a type II burst \cite{Gopalswamy10}, partly
because acceleration of electrons responsible for type II bursts may
depend on the shock conditions.  Based on \inlinecite{Nitta13a} and
confirmed in this study, LCPFs do not serve as a necessary condition
for the coronal shock wave because not all type II bursts are
associated with LCPFs.  However, when both of the phenomena are
observed in a given event, they provide information on when and where
the shock wave forms.  This is important for understanding the initial
evolution of SEP events \cite{Rouillard12}.  In a future work we will
derive thermal properties of the LCPFs from AIA multi-channel data to
understand how $v_{\tt LCPF}$ is related to the shock wave and
why Moreton--Ramsey waves are so rare (only two published and possibly two
more, A. Asai, 2014, private communication) in comparison with type II
bursts and LCPFs.


\section{Summary} 
      \label{S-Summary} 

We have studied the relation between LCPFs and type II bursts, by examining
the RSTN dynamic spectra for the periods of almost all of them during
April 2010\,--\,January 2013. 
Apart from the difficulty of isolating them objectively and the source
regions being far behind the limb, 
there are certainly type II bursts that do not accompany LCPFs,
notably when they are seen
primarily at high ($>$200~MHz) or low ($<$30~MHz) frequencies.   
When both phenomena are observed in a given eruptive event, 
they can be used to study when and where the shock wave forms.  
Based on AIA data in combination of STEREO EUVI and COR-1 data,
the arc-shaped propagating fronts can embrace the 3D structures of 
nascent CMEs rather than just their near-surface tracks, which have
usually been identified with EIT waves.  Those fronts moving toward
and beyond the limb are characteristic of type II bursts that start at 
low frequencies and high altitudes.  In these events, expansion in the
radial direction dominates, unlike other typical LCPFs where lateral
expansion is essential, and this may affect the location of the shock
waves that are responsible for type II bursts.  This idea may be
tested in future radio-imaging observations below 100~MHz.
Moreover, it may be important to distinguish CMEs or eruptions (not necessarily
detected by coronagraphs) that are initially dominated by radial or
lateral expansions for theory or models to contribute to reliable space
weather predictions.

\begin{acks}
This work has been supported by the NSF grant AGS-1259549, NASA AIA
contract NNG04EA00C and the NASA STEREO mission under NRL Contract
No. N00173-02-C-2035. NASA grant 
NNX11AO68G supported the work of WL.  
The work of NG and SY was supported by the NASA LWS TR\&T program.
\end{acks}

\bibliographystyle{spr-mp-sola}

\tracingmacros=2 \bibliography{typeII_lcpf_bibliography}
\IfFileExists{\jobname.bbl}{} {\typeout{}
\typeout{****************************************************}
\typeout{****************************************************}
\typeout{** Please run "bibtex \jobname" to obtain} \typeout{**
the bibliography and then re-run LaTeX} \typeout{** twice to fix
the references !}
\typeout{****************************************************}
\typeout{****************************************************}
\typeout{}}

\end{article} 

\end{document}